\documentclass{article}
\usepackage[margin=1in]{geometry}

\usepackage{graphicx} % Required for inserting images
\usepackage{authblk}  % For affiliations
\usepackage{orcidlink} % For ORCID icons, optional
\usepackage{makecell}
\usepackage{url}

\title{Texture Independently Drives Liking \\
in AI-Generated Alternative Protein Burgers}

\author[1]{Vahidullah Tac}
\author[1]{Aeneas O. Koosis}
\author[1]{Ellen Kuhl}

\affil[1]{Department of Mechanical Engineering, Stanford University, Stanford, California, United States}

\date{May 2026}

\begin{document}

\maketitle

\noindent\textbf{Correspondence:} ekuhl@stanford.edu

\begin{abstract}
Texture shapes how we perceive and like food, yet clear links between mechanical measurements and sensory perception of texture remain elusive. Here we combine sensory data from a blind tasting with 101 participants with mechanical texture profile analysis across six burgers to identify the textural features that drive consumer perception and liking. We compare five burgers--generated with artificial intelligence--with animal-based, plant-based, mushroom-based, and hybrid animal-mushroom patties, and the classical Big\,Mac. Three main findings emerge: First, animal-based burgers occupy a distinctive and coherent sensory-mechanical region associated with attributes such as firm, fatty, and holds together. Second, mushroom- and plant-based burgers deviate from this region in protein-dependent ways: mushroom-based burgers associate with springy and gummy textures, while plant-based burgers associate with dry, brittle, and crumbly textures. Hybrid animal-mushroom burgers, however, maintain sensory profiles comparable to fully animal-based burgers. Third, resilience emerges as the strongest mechanical correlate of perceived meatiness and sensory texture, while stiffness and hardness show no statistically significant association with consumer perception. Texture independently predicts overall liking alongside flavor: increasing texture liking by one point increases overall liking by 0.28. Among all sensory attributes, meatiness is the dominant predictor of texture liking. These findings identify resilience as a promising target for texture engineering and establish texture as a critical design objective for sustainable alternative proteins.
\end{abstract}


\section{Motivation}
Few foods have shaped human health and environmental sustainability as strongly as burgers. Burgers concentrate many of the defining challenges of modern food systems into a single product: nutrition, sustainability, flavor, texture, and consumer perception \cite{ilicMaterialsPropertiesOral2022}. Conventional meat production places a major burden on both human health and the environment \cite{Clark2022}. Animal agriculture contributes substantially to greenhouse gas emissions, land use, and water consumption \cite{Poore2018}. Diets rich in processed red meat also increase the risk of cardiovascular disease, metabolic disorders, and colorectal cancer \cite{Willett2019}. Plant- and mushroom-based proteins offer a promising path toward healthier and more sustainable foods \cite{Malila2024}. Yet consumer adoption remains limited by persistent gaps in flavor and texture relative to animal meat \cite{StPierre2024}. At the same time, generative artificial intelligence is emerging as a powerful tool for food design, capable of exploring large formulation spaces and optimizing recipes for nutrition, sustainability, and taste \cite{tacGenerativeArtificialIntelligence2026}. These advances create new opportunities to engineer alternative proteins with targeted sensory and mechanical properties \cite{tacGenerativeAIMaterial2026}.

Texture plays a central role in how we perceive and enjoy food \cite{Szczesniak2002}. In burgers, texture shapes sensory impressions such as softness, firmness, juiciness, fattiness, fibrousness, and meatiness, all of which strongly influence liking and purchase intent \cite{vandenBedem2026}. Texture exists on two fundamentally different levels \cite{StPierre2024}: Consumers perceive \textit{sensory texture} through chewing and oral processing, while laboratory instruments measure \textit{mechanical texture} through controlled deformation experiments \cite{Chen2009}. Establishing quantitative links between these two domains remains challenging, particularly for alternative proteins \cite{Dunne2025}. Animal meat derives its characteristic texture from the hierarchical architecture of muscle fibers, connective tissue, and fat networks \cite{Purslow2020}; plant- and mushroom-based meats instead form structurally different matrices with distinct deformation and fracture mechanisms \cite{Dekkers2018}. As a result, many plant- and mushroom-based meats fail to reproduce the cohesive, elastic, and meat-like texture of animal meat \cite{vervenneProbingMyceliumMechanics2025}. This makes alternative-protein texture engineering a heterogeneous soft-materials problem in which microstructure, phase connectivity, loading rate, and coupled deformation mechanisms govern the effective mechanical response  \cite{bahmaniKdtreeacceleratedHybridDatadriven2021a, upadhyayViscohyperelasticConstitutiveModeling2020}. Quantifying these mechanical responses is therefore essential for understanding why different burger formulations produce distinct sensory experiences.

Texture profile analysis provides a quantitative framework to characterize the mechanical texture of food through standardized and repeatable laboratory measurements \cite{Breene1975}. The method applies a double-compression test that mimics the process of chewing and extracts parameters such as stiffness, hardness, cohesiveness, springiness, resilience, and chewiness from the resulting force-displacement response \cite{Bourne1978}. Unlike subjective sensory evaluations, texture profile analysis generates reproducible measurements under controlled loading amplitudes and rates and enables direct comparison across formulations \cite{novakovicComparisonWarnerBratzlerShear2017}. However, the mechanical parameters from a texture profile analysis do not directly map onto human sensory perception, and it remains unclear which parameters best predict perceived texture and overall liking \cite{Dunne2025}. Here, we bridge this gap by integrating sensory surveys and texture profile analysis across six AI-designed burgers to characterize how mechanical texture shapes sensory perception and liking of animal-, plant-, mushroom-, and hybrid animal-mushroom burgers.

\begin{figure}[h]
\centering
\includegraphics[width=0.64\textwidth]{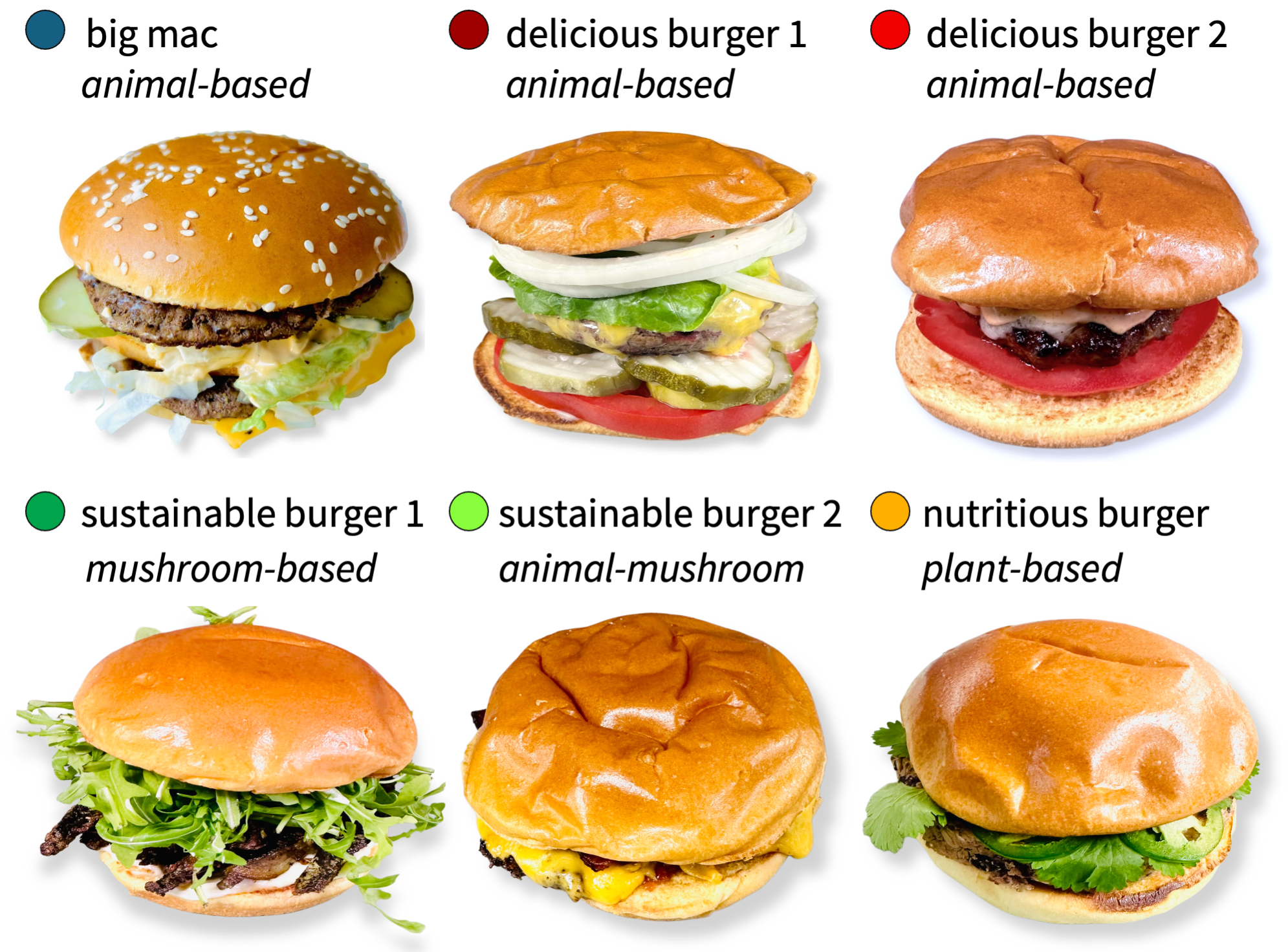} 
\caption{{\bf{Overview of the six burgers used in this study.}} 
We generated five new burgers using our generative artificial intelligence platform for burgers: 
delicious burger 1 and delicious burger 2 were optimized for palatability and contain animal meat; 
sustainable burger 1 was optimized for sustainability and uses mushroom; 
sustainable burger 2 uses a hybrid animal-mushroom blend; and 
nutritious burger uses beans. 
We use the classical Big\,Mac\textsuperscript{\textregistered} for benchmarking, it uses animal meat.}
\label{fig1}
\end{figure}

%%%%%%%%%%%%%%%%%%%%%%%%%%%%%%%%%%%%%%%%%%
\section{Materials and Methods}
We previously designed a generative artificial intelligence platform for food design \cite{tacGenerativeArtificialIntelligence2026}. 
%delicious burger 1, delicious burger 2, sustainable burger 1, sustainable burger 2, and nutritious burger. 
We used this tool to generate five burgers
and optimize 
for palatability in the animal-based delicious burgers 1 and 2, 
for environmental impact in the mushroom-based sustainable burger 1 
and the hybrid animal-mushroom sustainable burger 2, and 
for nutritional value in the plant-based nutritious burger (Figure \ref{fig1}). 

We conducted a sensory survey with $n =$ 101 respondents from the general population. Participants evaluated the textural properties, flavor, and overall liking of the five AI burgers and, as a benchmark comparison, the classical Big\,Mac\textsuperscript{\textregistered}. Here, we use the survey data to assess the effect of patty type on the perceived texture, investigate which textural attributes contribute most heavily to the participant's liking of a burger, and explore how animal- and plant-based burgers compare along those attributes. 

We also perform texture profile analysis tests \cite{novakovicComparisonWarnerBratzlerShear2017, st.pierreMimickingMechanicsComparison2024a, vervenneProbingMyceliumMechanics2025} of the six patties to quantify the texture of the burgers in the laboratory. We obtain texture profile analysis parameters such as stiffness, hardness and cohesiveness for each of the six burgers and investigate whether these mechanical attributes can be used to detect perceived textural attributes.

\subsection{Generative AI for burgers}
Machine learning models and generative AI are increasingly used to learn compact, lower-dimensional representations of complex systems and design spaces; ranging from food safety to physical systems like materials modeling and sediment transport \cite{qianHowCanAI2023, amiri-hezavehPhysicsAugmentedMachineLearning2025, nazariReducedorderModelingTemporal2025, thomasMachineLearningApproach2022}. We use generative AI to learn human preferences in food, specifically burgers, and use the resulting model for targeted recipe generation for goals such as environmental sustainability and nutritional value. 
We represent each burger recipe in terms of the ingredient selection and ingredient quantification. Thus, burger recipes are hybrid discrete-continuous objects and modeling such data requires a two-stage approach that respects the hybrid nature of the data. We create a generative AI model that combines a discrete diffusion model and a continuous diffusion model to generate new recipes for burgers \cite{tacGenerativeArtificialIntelligence2026,tacGenerativeAIMaterial2026}. We denote the two sub-models as the \emph{mask model} and the \emph{weight model}. 

The \emph{mask model} uses multinomial diffusion to sample discrete data. Multinomial diffusion uses categorical distributions for learning probability distributions over complex and high dimensional discrete data \cite{NEURIPS2021_67d96d45}. In the case of ingredient selection for burger recipes, there are only two categories; absent (category 0) or present (category 1). The mask model thus generates a binary mask, $\vec{m}$, indicating which ingredients will be used in a given recipe. We then use this mask in the weight model for determining ingredient weights.

The \emph{weight model} uses a continuous score-based diffusion model to sample the weights of the ingredients selected by the mask model. Score-based diffusion models use stochastic differential equations to noise and denoise the data, where the denoising step relies on a neural-network based approximation of the score function of the target probability distribution \cite{tacGenerativeHyperelasticityPhysicsinformed2024, songScoreBasedGenerativeModeling2021}. In the case of burger recipes, the score function is given as $\nabla_{\vec{w}} \log p_t(\vec{w}|\vec{m})$ where $\vec{w}$ is the vector of weights and $\vec{m}$ is the binary mask of the recipe. 

We train the generative model on a custom dataset of more than 2,260 human designed recipes curated from \verb|Food.com| \cite{tacGenerativeArtificialIntelligence2026}.

\subsection{Sensory survey}
We conduct a blind sensory survey with $n=101$ participants from the general population in an active restaurant in San Francisco, California. 
Each participant evaluates all five AI-designed burgers and the Big\,Mac\textsuperscript{\textregistered} on a 7-point Likert scale for \emph{overall liking}, \emph{flavor}, and \emph{texture} \cite{stpierre2024_first} and on a 5-point Likert scale for five texture-related attributes, and answers a check-all-that-apply question that involves 15 texture-related attributes. We treat the Likert-scale responses as continuous variables, consistent with common practice in sensory research \cite{kooLikertTypeScale2025}. The texture-related questions focus on the \emph{patty} in the burger, while the remaining questions focus on the entire \emph{burger}. Table \ref{tab:survey_questions} summarizes the full list of questions. 

All responses are fully anonymized. The serving order is randomized per participant. We do not relay information about the burger such as its ingredients or its name to the participants. All burgers, except the Big\,Mac\textsuperscript{\textregistered}, were prepared in the kitchen of the restaurant by the same group of chefs following the same recipes and served fresh. The Big\,Mac\textsuperscript{\textregistered} was purchased from McDonald's and kept fresh on a heated tray. All burgers are served in the same session. Participants were offered water as a palate cleanser between burgers. 
This project was determined by the Stanford University Institutional Review Board to not meet the definition of human subjects research as defined in 45 CFR 46.102.

\begin{table}[t]
\centering
\caption{{\textbf{Sensory survey questions.}} Attributes, question types, question statements, and choices.}
\label{tab:survey_questions}
\begin{tabular}{|l|l|l|l|}  \hline
        \textbf{attribute} & \textbf{type} & \textbf{question} & \textbf{choices}\\
        \hline \hline
        \textbf{overall liking} & Likert & \makecell[l]{How would you rate your \\ overall liking of the burger?} & 1--7\\ \hline
        \textbf{flavor liking}  & Likert   & \makecell[l]{How would you rate the flavor \\ of the burger?} & 1--7 \\ \hline
        \textbf{texture liking} & Likert & \makecell[l]{How would you rate the texture \\ of the burger?} & 1--7 \\ \hline \hline
        \makecell[l]{\textbf{sensory} \\ \textbf{attributes}} & CATA & \makecell[l]{Please check all words or phrases  that \\  describe the texture of the burger.} & chewy, sticky, $\ldots$ \\ \hline \hline
        \textbf{softness} & Likert & How soft is the patty in the burger? & 1--5 \\ \hline
        \textbf{hardness} & Likert & How hard is the patty in the burger? & 1--5 \\ \hline
        \textbf{fattiness} & Likert & How fatty is the patty in the burger? & 1--5 \\ \hline
        \textbf{moistness} & Likert & How moist is the patty in the burger? & 1--5 \\ \hline
        \textbf{fibrousness} & Likert & How fibrous is the patty in the burger? & 1--5 \\ \hline
\end{tabular}
\end{table}

\subsection{Texture profile analysis}

Texture profile analysis is a popular method for characterizing the mechanical properties of foods in laboratory settings \cite{novakovicComparisonWarnerBratzlerShear2017, st.pierreMimickingMechanicsComparison2024a, vervenneProbingMyceliumMechanics2025}. 
\begin{figure}[h]
\centering
\includegraphics[width=0.78\textwidth]{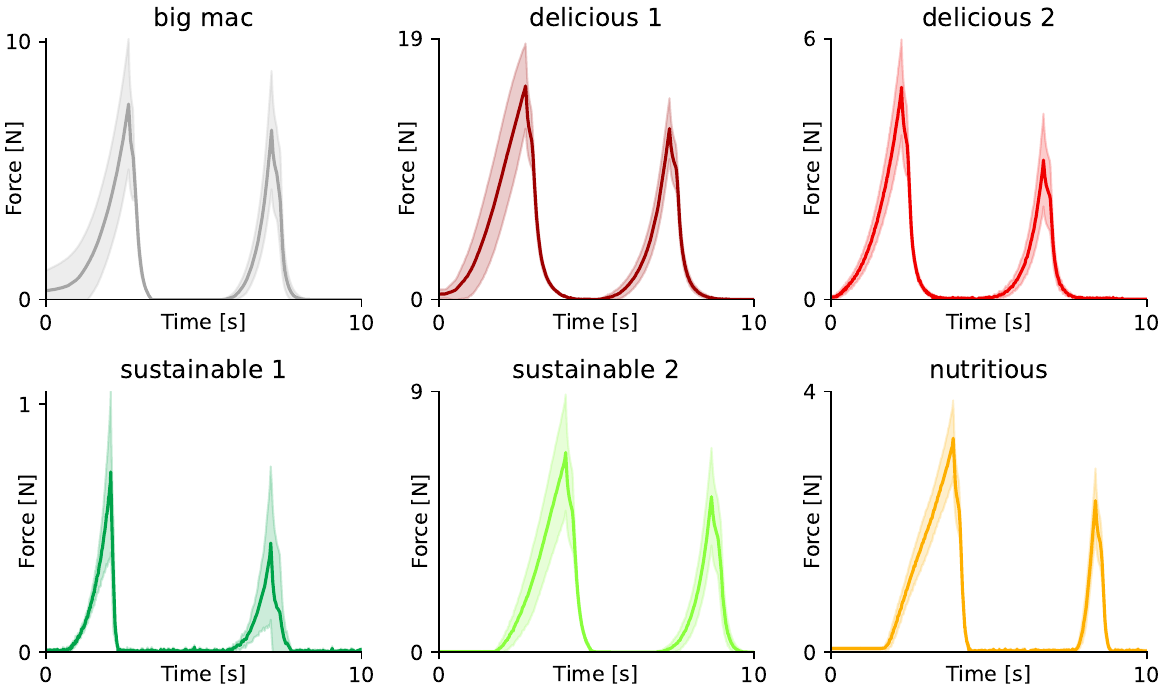} 
\caption{{\textbf{Texture profile analysis results for the six burger patties.}} 
The lines and shaded areas represent ensemble means and 95\% confidence intervals for the force versus time responses of $n=10$ samples from each patty under the double-compression test. 
The narrow confidence intervals indicate that the tests are highly reproducible despite the material heterogeneity of the patties.}
\label{fig2}
\end{figure}
It typically involves double-compression tests on a rheometer at loading rates that resemble typical chewing speeds \cite{ilicMaterialsPropertiesOral2022, novakovicComparisonWarnerBratzlerShear2017}. 
Within this study, we performed texture profile analysis within one hour of cooking to capture the mechanical properties of fresh patties.
We collect data on $n=$10 samples from each patty, 
where the samples are prepared with an 8 mm biopsy punch. 
We perform two compression cycles to $-50\%$ strain at a rate of $-25\%/s$ and calculate the ensemble means and 95\% confidence intervals of the force versus time data across  $n=$10 samples for all six patties (Fig. \ref{fig2}). From the mean curves, we calculate six texture profile analysis parameters including stiffness, hardness, cohesiveness, springiness, resilience and chewiness \cite{st.pierreMimickingMechanicsComparison2024a}. 

\subsection{Statistical analysis}
\subsubsection{Correspondence analysis}
We use correspondence analysis (CA) \cite{abdiCorrespondenceAnalysis2017} to examine relationships among burger samples, sensory texture attributes collected through the check-all-that-apply (CATA) question, and mechanical texture profile analysis parameters. We construct a contingency table $\mathbf{N} = (n_{ij})$, where rows represent burger samples and columns represent CATA attributes, with entries equal to the total of attribute selection for each sample. We convert this table into relative frequencies $\mathbf{P} = \mathbf{N}/n$, where $n = \sum_{i,j} n_{ij}$. We compute row and column marginal proportions as $r_i = \sum_j p_{ij}$ and $c_j = \sum_i p_{ij}$ and center the data with respect to the independence model,
$ \mathbf{Z} 
= \mathbf{D}_r^{-1/2} [\, \mathbf{P} - \mathbf{r}\mathbf{c}^{\rm{t}} \,] \mathbf{D}_c^{-1/2}$,
where $\mathbf{D}_r$ and $\mathbf{D}_c$ denote diagonal matrices of row and column masses. We apply a singular value decomposition to $\mathbf{Z}$ to obtain principal axes and singular values, and define eigenvalues as $\lambda_k = \sigma_k^2$ to quantify the inertia explained by each dimension.

We incorporate the texture profile analysis parameters as supplementary variables and project them onto the correspondence analysis space without affecting the factorization. We calculate correlations between each texture profile analysis variable and the principal dimensions and represent these variables as vectors based on their loadings. We compute principal coordinates for burger samples and sensory attributes and display them in a common low-dimensional space. We measure distances between row profiles using the chi-square metric,
$d^2(i,i') 
= \sum_j  
( {p_{ij}}/{r_i} - {p_{i'j}}/{r_{i'}})^2 / {c_j}$.
In this representation, proximity between samples and attributes indicates association, while larger distances reflect differences in sensory profiles relative to expectations under independence. This approach enables joint interpretation of consumer-perceived texture attributes and mechanical texture measurements across burger formulations.
We use the \verb|prince| library in Python for the implementation of the correspondence analysis \cite{Halford_Prince}.

\subsubsection{Linear correlations}
We evaluate linear relationships between sensory texture attributes and mechanical texture profile analysis parameters using Pearson correlation coefficients. For each pair consisting of a CATA attribute (aggregated across survey respondents) and a texture profile analysis parameter, we calculate the Pearson correlation coefficient (r) and the associated p-value. When $p < 0.05$, we classify the relationship as statistically significant and interpret the direction of association based on the sign of $r$, where positive values indicate positive correlations and negative values indicate negative correlations. We explicitly show the plots of the statistically significant correlations in the figures and label them according to their direction and magnitude. As a result of aggregation of CATA attributes, we perform this analysis on the burger level, $n=6$. Given this small number of samples, we interpret the correlations that we find through this method as exploratory.

\subsubsection{Mixed effects modeling}
We implement a linear mixed-effects model to assess the relative importance of flavor and texture on overall liking while accounting for repeated measurements within participants \cite{gelman_hill_2006}. The n = 101 participants rated the six burgers in overall liking, flavor, and texture, resulting in 606 observations. The model treats overall liking as the dependent variable, flavor and texture scores as fixed-effect predictors, and includes a random intercept for each participant to account for individual baseline differences. We standardize all scores as z-scores to allow direct comparison of effect sizes. This results in the following mathematical model,
$z_{ij} = \beta_0 + \beta_1 \cdot x_{ij} + \beta_2 \cdot y_{ij} + u_i + \epsilon_{ij}$,
where $z_{ij}, \, x_{ij}, \, y_{ij}$ denote the overall liking, flavor, and texture scores for participant $i$ on burger $j$,
$\beta_0$ is the fixed-effect intercept,
$\beta_1$ and $\beta_2$ are the fixed-effect coefficients for flavor and texture,
$u_i \sim \mathcal{N}(0, \sigma_u^2)$ is the participant-specific random intercept to capture baseline differences between participants, and
$\epsilon_{ij} \sim \mathcal{N}(0, \sigma^2)$ is the residual error.
We use the \verb|mixedlm| class from the \verb|statsmodels| package in Python for the implementation of the mixed effects model \cite{seabold2010statsmodels}.

%%%%%%%%%%%%%%%%%%%%%%%%%%%%%%%%%%%%%%%%%%
\section{Results}

\begin{figure}[h]
\centering
\includegraphics[width=\textwidth]{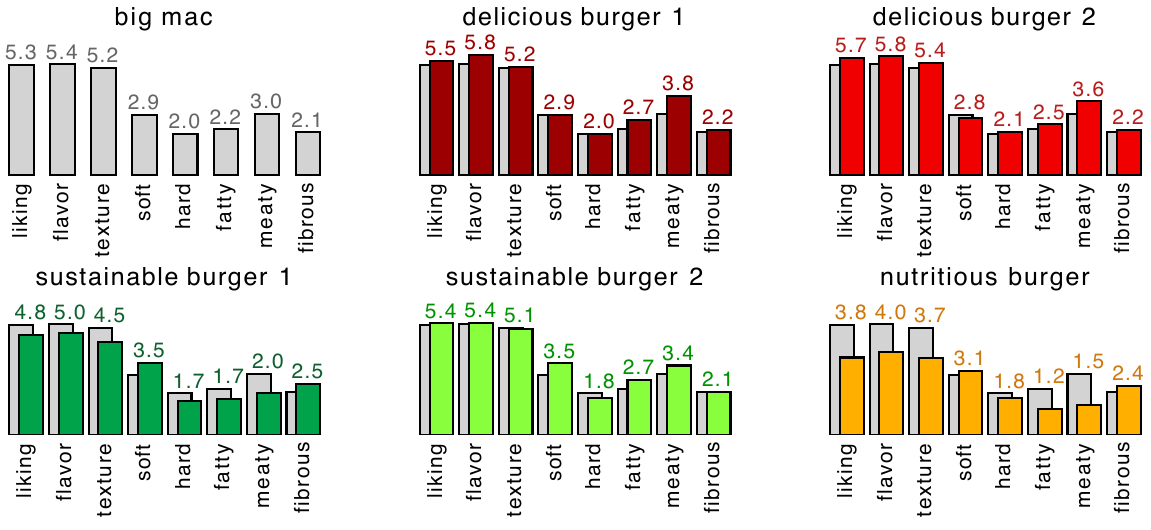} 
\caption{{\textbf{Mean sensory ratings for overall liking, flavor, texture, and five texture-related attributes.}} 
The plant-based burgers, sustainable burger 1 and nutritious burger, score lower on overall liking, flavor, texture, hardness, fattiness, and meatiness, 
but score higher on softness and fibrousness, compared to the animal-based burgers.}
\label{fig3}
\end{figure}

Sensory evaluation across 101 participants reveals clear differences between animal- and  plant-based burgers along most dimensions (Figure \ref{fig3}). The two AI-optimized delicious burgers outperform the Big Mac® on overall liking (5.5 and 5.7 vs. 5.3), suggesting that generative AI can successfully optimize for palatability beyond what an established commercial product achieves. Among all burgers, the nutritious burger scores lowest across all three primary categories — liking (3.8), flavor (4.0), and texture (3.7) — which indicates that nutritional optimization comes at a significant cost to consumer acceptance. The plant-based burgers score higher on softness but lower on hardness, fattiness, and meatiness compared to the animal-based burgers, which points to clear textural differences between the two groups. Notably, the sustainable burger 2 that uses an animal-meat-mushroom blend, scores comparably to the entirely animal-based delicious burgers on most attributes, suggesting that partial substitution of animal protein with mushroom does not substantially degrade texture perception.

%\begin{figure}[h]
%\centering
%\includegraphics[width=\textwidth]{fig4} 
%\caption{{\textbf{Comparison of texture profile analysis parameters and sensory texture ratings.}} Mechanical properties span wide ranges across all six burgers, while sensory ratings remain compressed. The order of burgers along the mechanical parameters does not always mirror the order along sensory ratings.}
%\label{fig4}
%\end{figure}

\begin{figure}[h]
        \centering
        \includegraphics[width=\textwidth]{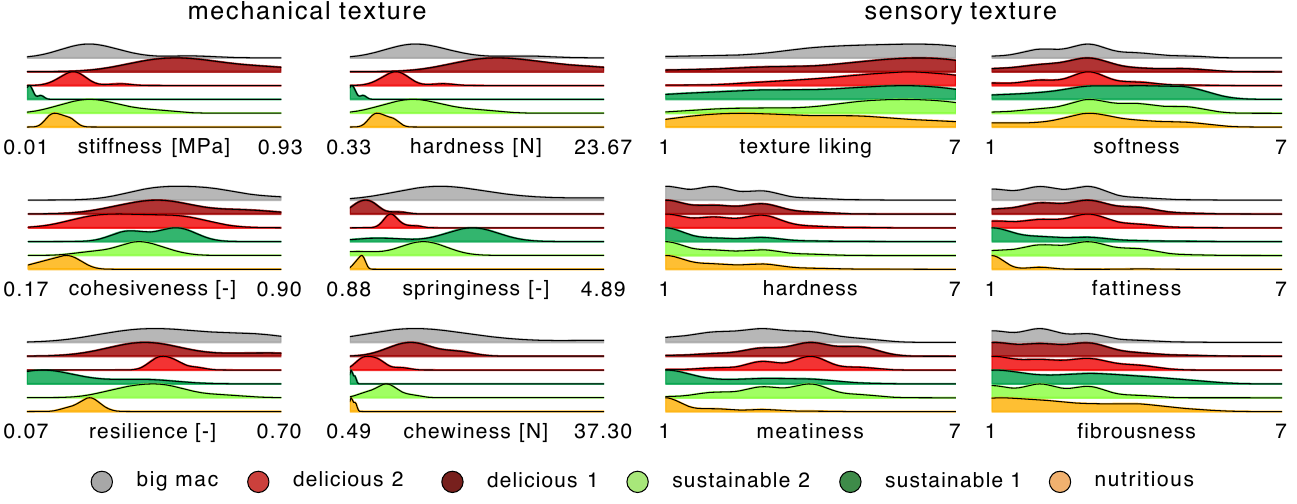}
        \caption{{\textbf{Comparison of texture profile analysis parameters and sensory texture ratings.}} Mechanical properties span wide ranges across all six burgers, while sensory ratings remain compressed. The order of burgers along the mechanical parameters does not always mirror the order along sensory ratings.}
\label{fig4}
\end{figure}

Comparing the mechanical and sensory texture data side by side reveals a striking asymmetry in the dynamic range (Figure \ref{fig4}). Texture profile analysis parameters span wide absolute ranges across all burgers — hardness varies from 0.33 N to 23.67 N and chewiness from 0.49 N to 37.30 N — whereas sensory ratings on the 7-point Likert scale remain compressed within a narrow band for most attributes. This compression in the sensory space suggests that human perception of texture is nonlinear and likely subject to adaptation and contrast effects; large mechanical differences between burgers do not translate proportionally into perceived textural differences. Notably, the ordering of burgers along the mechanical parameters does not always mirror their ordering in sensory ratings. This suggests that certain texture profile analysis parameters may be poor proxies for perceived texture. These observations motivate a more systematic investigation to characterize which specific mechanical parameters actually drive sensory perception, as pursued in the subsequent analyses.

\begin{figure}[h]
\centering
\includegraphics[width=0.8\textwidth]{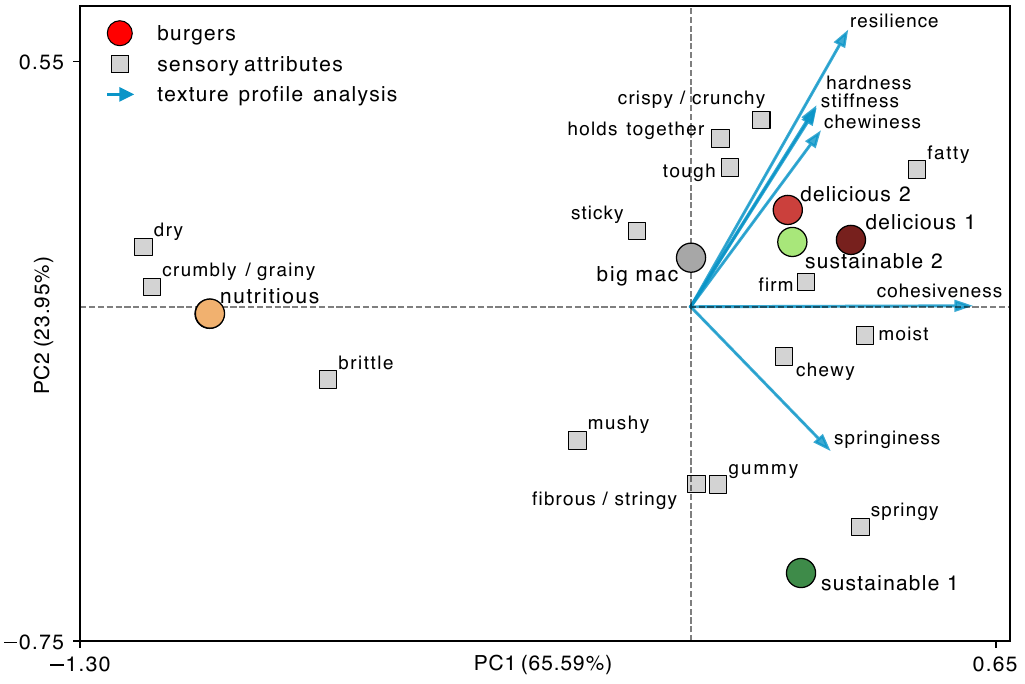} 
\caption{{\textbf{Correspondence analysis biplot that relates burgers, sensory texture attributes, and texture profile analysis parameters in a shared low-dimensional space.}} Meat burgers form a cluster in the upper right quadrant along with sensory attributes such as firm, holds together, and fatty, and most texture profile analysis parameters. The plant-based nutritious burger has a lower value on principal component 1 (PC1), and is associated with attributes such as dry, crumbly, and brittle, whereas the mushroom-based sustainable burger 1 scores lower on principal component 2 (PC2) and is associated with attributes like springy and gummy.}
\label{fig5}
\end{figure}

The correspondence analysis biplot reveals that the principal components, PC1 (65.59\% of inertia) and PC2 (23.95\%), jointly separate burgers by patty protein source (Figure \ref{fig5}). Animal-based burgers (Big Mac®, delicious burger 1, delicious burger 2, sustainable burger 2) cluster in the upper right of the space, scoring high on both PC1 and PC2, and co-locate with sensory attributes such as firm, holds together, fatty, and tough, as well as most texture profile analysis parameters. The two alternative-protein burgers deviate from this animal-based cluster in distinctly different directions: the plant-based nutritious burger scores low on PC1 and associates with attributes such as dry, crumbly/grainy, and brittle, while the mushroom-based sustainable burger 1 scores low on PC2 and associates with attributes like springy and gummy. This orthogonal divergence indicates that the two alternative proteins produce qualitatively distinct textural profiles relative to animal meat. The bean patty deviates primarily along the dimension that separates firm, cohesive textures from dry and brittle ones, while the mushroom patty deviates along the dimension that distinguishes elastic, springy textures from the tough and fatty character of meat.

\begin{figure}[h]
\centering
\includegraphics[width=\textwidth]{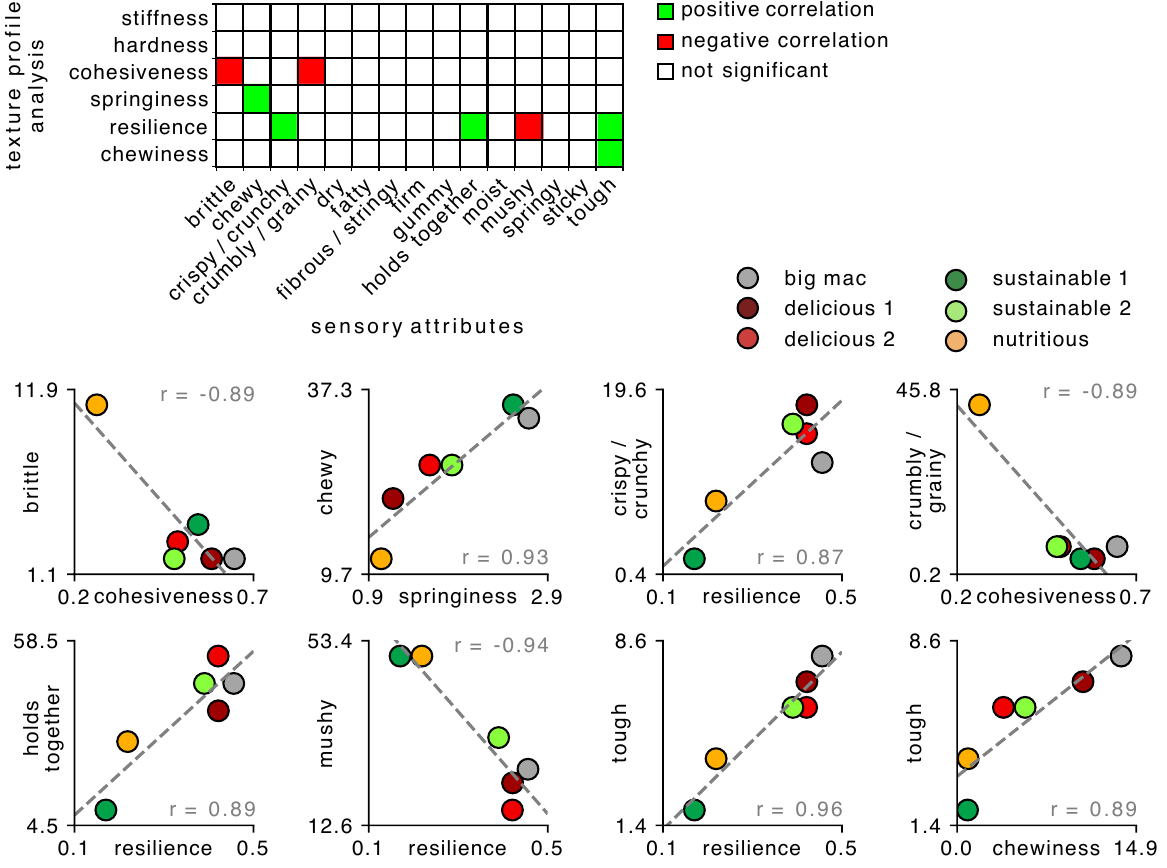} 
\caption{\textbf{Linear correlations between texture profile analysis parameters and sensory attributes.} The data suggest that resilience is particularly promising for texture engineering, as it has significant correlations with four sensory attributes; crispy/crunchy, holds together, mushy and tough. Cohesiveness, springiness and chewiness also have significant correlations with sensory attributes. However stiffness and hardness have no statistically significant correlations.}
\label{fig6}
\end{figure}

Among the six texture profile analysis parameters, resilience emerges as the most informationally rich predictor of consumer-perceived texture; it yields statistically significant correlations with four distinct sensory attributes: crispy/crunchy ($r = 0.87$), holds together ($r = 0.89$), mushy ($r = -0.94$), and tough ($r = 0.96$) (Figure \ref{fig6}). The breadth and magnitude of these correlations suggest that resilience, a measure of how readily a material recovers energy after deformation, captures something fundamental about the structural integrity of a patty as perceived during chewing. Cohesiveness similarly shows strong negative correlations with brittle ($r = -0.89$) and crumbly/grainy ($r = -0.89$), which reinforces the interpretation that particle cohesion in the patty matrix underlies multiple related textural impressions. In contrast, stiffness and hardness, arguably the most intuitive mechanical descriptors, show no statistically significant correlations with any sensory attribute, a counterintuitive result that underscores the importance of empirical validation over assumptions when linking mechanical measurements to human perception. 

\begin{figure}[h]
\centering
\includegraphics[width=\textwidth]{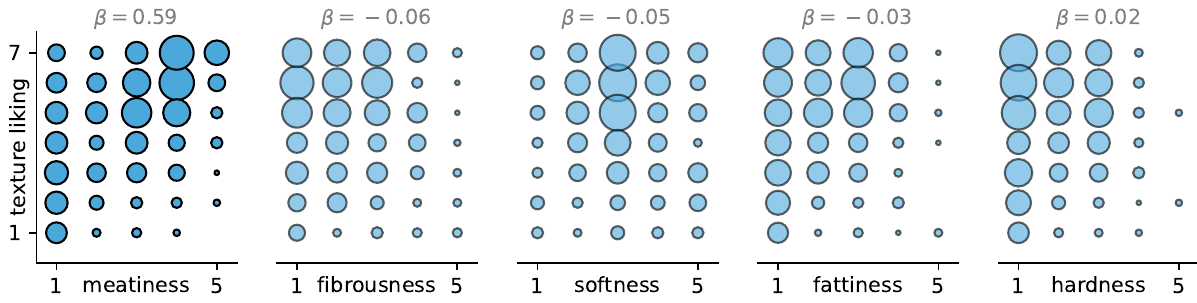} 
\caption{\textbf{Correlation between texture liking and textural attributes.} 
Texture liking is rated on a 7-point Likert scale, 
textural attributes are rated on a 5-point Likert scale. 
Meatiness shows the strongest association with perceived texture quality with a 1-point increase in meatiness independently explaining 0.59 point increase in texture liking in linear mixed-effects modeling.}
\label{fig7}
\end{figure}

Of the five textural attributes probed with 5-point Likert questions, meatiness is by far the dominant predictor of texture liking, with a mixed-effects coefficient of $\beta = 0.59$, meaning that a 1-point increase in perceived meatiness independently predicts a 0.59-point increase in texture liking (Figure \ref{fig7}). The remaining attributes, fibrousness ($\beta = -0.06$), softness ($\beta = -0.05$), fattiness ($\beta = 0.03$), and hardness ($\beta = 0.02$), have coefficients near zero, indicating that they contribute negligibly to texture liking after accounting for meatiness and participant-level random effects. This finding reframes the challenge of meat analog design: the goal should not simply be to match any individual textural property in isolation, but specifically to achieve perceived meatiness, which appears to serve as a holistic proxy for the constellation of textural cues that consumers associate with a satisfying burger patty.

\begin{figure}[h]
\centering
\includegraphics[width=0.75\textwidth]{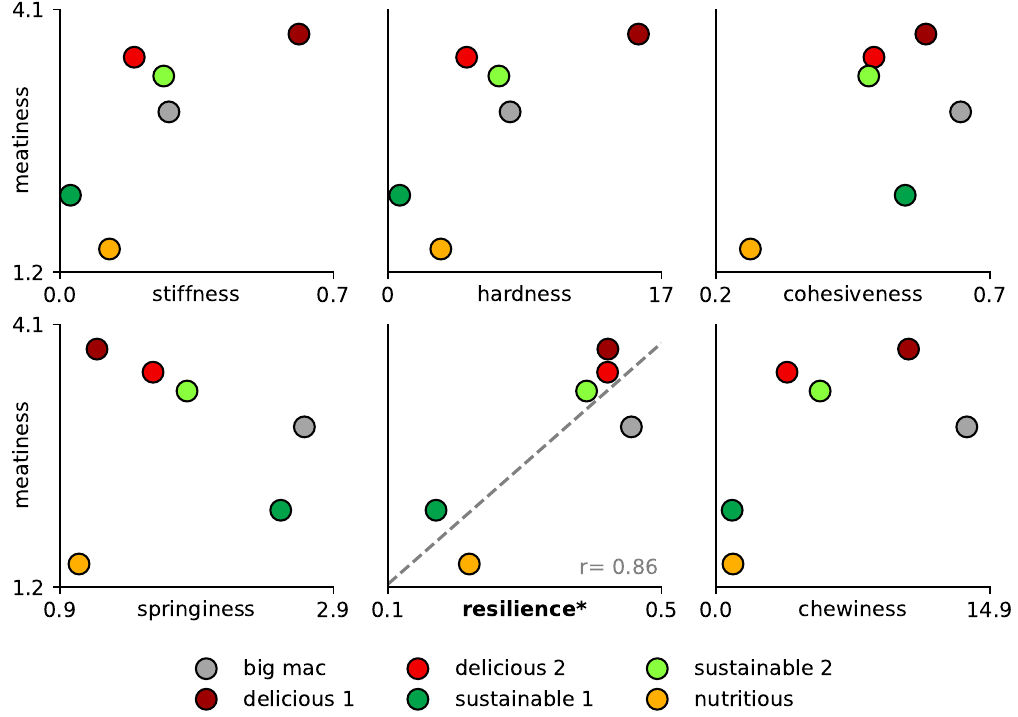} 
\caption{\textbf{Relationship between texture profile analysis parameters and perceived meatiness.} Resilience shows a strong positive correlation with meatiness, while stiffness, hardness, cohesiveness, springiness, and chewiness show no statistically significant association.}
\label{fig8}
\end{figure}

Examining the relationship between individual texture profile analysis parameters and perceived meatiness confirms that resilience is the strongest mechanical correlate of meat-like texture ($r = 0.92$), while stiffness, hardness, cohesiveness, springiness, and chewiness show no significant association (Figure \ref{fig8}). In view of the findings that meatiness is the primary driver of texture liking (Figure \ref{fig7}) and that resilience is the richest correlate of sensory attributes (Figure \ref{fig6}), resilience now emerges as a key mechanical target for meat analog engineering. Physically, a high-resilience material rapidly recovers its shape after compression, which mimics the elastic snap-back characteristic of muscle fiber networks in animal meat. The absence of significant correlations to stiffness and hardness is particularly notable: a patty can be made stiffer or harder through various means without becoming more meat-like in the eyes of the consumer. This illustrates the pitfall of targeting readily measurable but perceptually irrelevant mechanical properties.

\begin{figure}[h]
\centering
\includegraphics[width=0.78\textwidth]{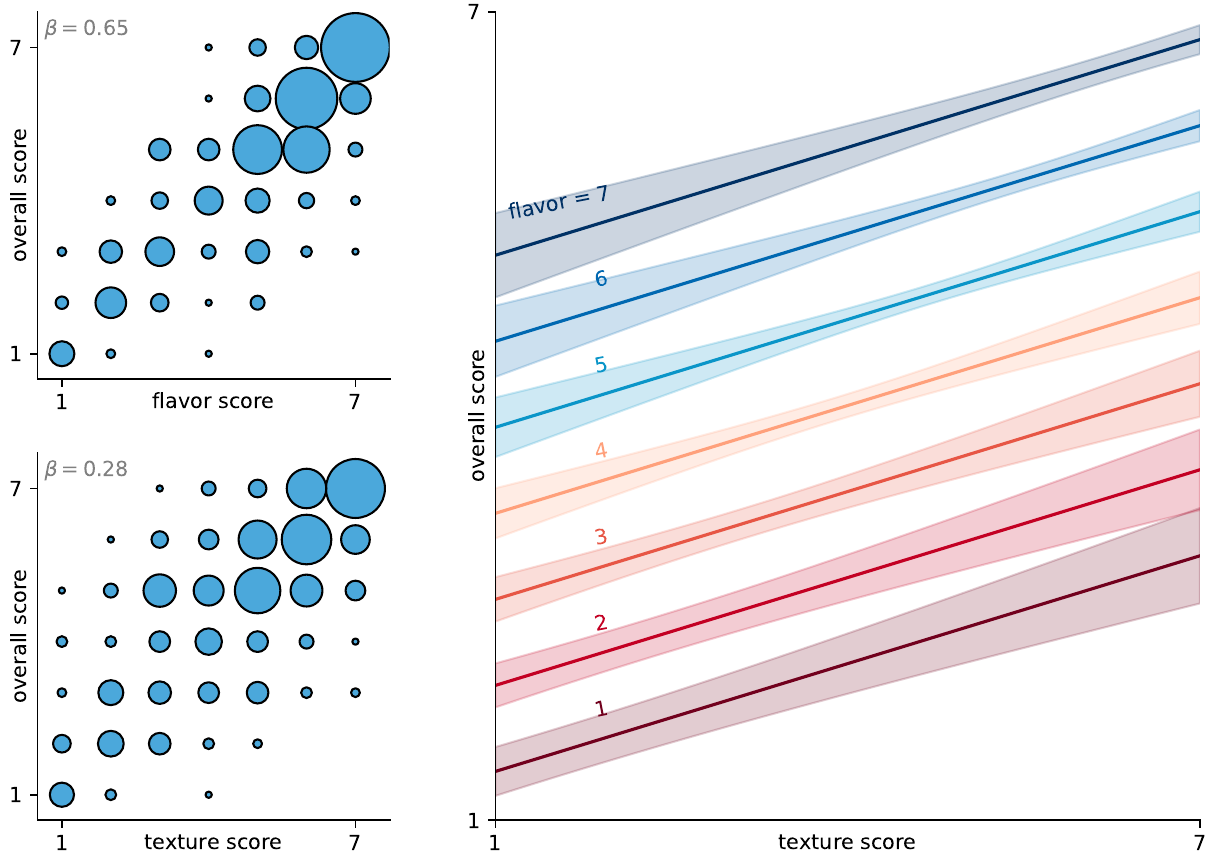} 
\caption{\textbf{Mixed-effects model shows the relationship between overall liking, flavor, and texture.} Texture has a strong impact in determining overall liking. For the same flavor score, a 1-point increase in texture score results in a 0.28-point increase in overall score, on average.}
\label{fig9}
\end{figure}

The mixed-effects model reveals that both flavor ($\beta = 0.65$) and texture ($\beta = 0.28$) are independently significant predictors of overall burger liking (Figure \ref{fig9}). While flavor is the stronger driver, texture makes a substantial independent contribution that cannot be dismissed, a result that justifies dedicated texture optimization in burger design alongside flavor. The right panel illustrates the practical magnitude of this texture effect: along the flavor = 7 axis (blue), texture can take a burger from a somewhat above-average experience (overall score 5.19 at the low-texture end) to an excellent one (6.79 at the high-texture end). The same dynamic plays out at the other extreme; along the flavor = 1 axis (red), good texture is the differentiator between a completely unpalatable burger (1.36) and one that is at least marginally acceptable (2.96). Perhaps most strikingly, even a burger with perfect flavor never reaches its full potential without strong texture: the flavor = 7 axis is still below 5.5 in overall liking when the texture score is 1, even after accounting for uncertainty in the predicted mean. This underscores that flavor excellence alone is insufficient for a truly outstanding consumer experience.

%%%%%%%%%%%%%%%%%%%%%%%%%%%%%%%%%%%%%%%%%%
\section{Discussion}
This study integrates sensory surveys and texture profile analysis to investigate how texture shapes perception and liking in AI-designed animal-, plant-, and mushroom-based burgers. Although the study focuses on a small set of burgers and broader validation across additional formulations remains necessary, three overarching findings emerge from this work:

The first finding is that  animal-based patties carry a distinctive {\it{textural signature}} that consumers reliably identify. In the correspondence analysis, animal-based burgers cluster tightly together in a shared sensory-mechanical space, co-locating with attributes such as firm, holds together, and fatty, and with most texture profile analysis parameters. This coherence across both sensory and mechanical measurements suggests that the texture of animal meat is not merely one point in a continuous textural space, but a recognizable and relatively compact region that consumers have strong and consistent associations with. 

The second finding is that plant- and mushroom-based patties deviate from this animal-based textural region in protein-dependent ways. The plant-based nutritious burger associates with dry, crumbly, and brittle attributes, while the mushroom-based sustainable burger 1 associates with springy and gummy ones, two qualitatively distinct failure modes that occupy different regions of the sensory space. This divergence has practical consequences for meat analog development: it suggests that there is no single universal strategy for {\it{closing the texture gap}} with animal meat, and that engineering interventions must be tailored to the specific protein matrix in use. The partial substitution of animal protein with mushroom in sustainable burger 2, however, does not substantially degrade perceived texture relative to fully meat-based burgers, pointing to a potentially viable middle path for reducing the environmental footprint of burgers without a large penalty in consumer acceptance.

The third finding concerns the specific mechanical properties that underlie meat-like texture perception. Among the six texture profile analysis parameters, {\it{resilience}} emerges as the most informative. It correlates significantly with four sensory attributes and shows the strongest association with perceived {\it{meatiness}} ($r = 0.92$). Since meatiness is itself the dominant predictor of texture liking ($\beta = 0.59$), resilience effectively connects laboratory measurements to consumer acceptance through a two-step chain: high resilience predicts high perceived meatiness, and high perceived meatiness predicts high texture liking. Stiffness and hardness, by contrast, show no statistically significant correlations with any sensory attribute, a counterintuitive result that cautions against targeting the most readily measurable mechanical properties without empirical validation of their perceptual relevance. For meat analog developers, this points to resilience as a more actionable engineering target than hardness or stiffness.

Finally, the mixed-effects model establishes that texture is an independent driver of overall consumer liking, with a 1-point increase in texture score predicting a 0.28-point increase in overall liking regardless of flavor. While flavor remains the stronger predictor, the contribution of texture is large enough to meaningfully shift consumer perception across the full range of the rating scale. A burger with excellent flavor but poor texture still falls short of its potential, while strong texture can elevate an average burger into an excellent one. For the development of sustainable and nutritious meat analogs, this finding is consequential: achieving flavor parity with animal meat is a necessary but insufficient condition for market success, and texture must be engineered with equal intentionality.

%%%%%%%%%%%%%%%%%%%%%%%%%%%%%%%%%%%%%%%%%%
\section{Conclusions}
This study establishes texture as an independent and consequential driver of consumer liking in animal-, plant-, and mushroom-based burgers, alongside flavor. By integrating a mechanical texture profile analysis with a sensory survey of 101 participants on six burger formulations, we show that animal meat patties occupy a distinctive and coherent region of the textural space that consumers reliably recognize, and that plant- and mushroom-based patties deviate from this region in protein-dependent ways. Among the mechanical properties we examine, resilience stands out as the most promising target for texture engineering, as it correlates strongly with perceived meatiness, the dominant predictor of texture liking. These findings suggest a concrete path forward for meat analog development: rather than targeting hardness or stiffness, which show no significant perceptual correlates, developers should prioritize the elastic recovery characteristics of the patty matrix. More broadly, the results reinforce that flavor optimization alone is insufficient for consumer acceptance of novel burger formulations; texture must receive equal engineering attention if nutritionally or environmentally optimized burgers are to close the gap with their traditional animal meat counterparts.

\section*{Funding}

This research was supported by
the Schmidt Science Fellowship 
in partnership with the Rhodes Trust to Vahidullah Tac, and by
the Stanford Bio-X Snack Grant 2025,
the Stanford SDSS Accelerator Grant 2025,
the NSF CMMI Award 2320933, and
the ERC Advanced Grant 101141626 to Ellen Kuhl.

\section*{Data availability}
Our source code, data, and examples are available at 
https:/\!/github.com/LivingMatterLab/AI4Food.

\section*{Acknowledgements}
We thank 
Executive Chef Justin Schneider 
for his culinary expertise in creating preparation instructions, 
Caroline Cotto from NECTAR at Food System Innovations 
for stimulating discussions,
and
Alice Wistar and Alex Weissman from Palate Insights
for performing the consumer survey.
We acknowledge 
access to the Stanford Marlowe Computing Platform for high performance computing.

\bibliographystyle{unsrt}
\bibliography{references}

\end{document}